\documentclass[11pt,oneside,letter]{article}
\usepackage[papersize={8.5in,11in}]{geometry}
\usepackage{amsmath,amssymb,enumitem,accents}
\usepackage{dsfont,mathrsfs}
\usepackage{hyperref,appendix}
\usepackage{bm}
\usepackage[hang,flushmargin]{footmisc}
\usepackage{graphicx}

\makeindex
\usepackage[square,comma,numbers,sort&compress]{natbib}
\usepackage{xcolor}
\definecolor{dark-red}{rgb}{0.5,0.1,0.1}
\definecolor{dark-blue}{rgb}{0.1,0.1,0.5}
\definecolor{medium-blue}{rgb}{0,0,0.6}
\hypersetup{
  pdfborder={0 0 0},
  colorlinks, linkcolor={dark-red},
  citecolor={dark-blue}, urlcolor={medium-blue}
}

\newcommand{\qed}{\nobreak \ifvmode \relax \else
      \ifdim\lastskip<1.5em \hskip-\lastskip
      \hskip1.5em plus0em minus0.5em \fi \nobreak
      \vrule height0.75em width0.5em depth0.25em\fi}

\def\bqt{\begin{quotation}}   \def\eqt{\end{quotation}}
\def\beq{\begin{equation}}   \def\eeq{\end{equation}}
\def\bea{\begin{eqnarray}}  \def\eea{\end{eqnarray}}
\def\eqnarraynum{\hfill\parbox{.4in}{\begin{eqnarray}\end{eqnarray}}}

\def\noi{\noindent}
\newcounter{saveeqn}%
\newcommand{\alphaeqn}{\setcounter{saveeqn}{\value{equation}}%
\stepcounter{saveeqn}\setcounter{equation}{0}%
\renewcommand{\theequation}
 {\mbox{\arabic{saveeqn}-\alph{equation}}}}%
\newcommand{\reseteqn}{\setcounter{equation}{\value{saveeqn}}
\renewcommand{\theequation}{\arabic{equation}}}

\def\blist{\begin{list}{$\bullet$}{\setlength{\leftmargin}{.15in}
      \setlength{\itemsep}{0pt} \setlength{\topsep}{0pt}
      \setlength{\parsep}{0pt}}}

\def\B{$\mathscr{B}$}
\def\mB{\mathscr{B}}

\def\mG{\mathcal{G}}
\def\M{$\mathscr{M}$}
\def\mM{\mathscr{M}}

\def\T{$\mathcal{T}$}
\def\mT{\mathcal{T}}
\def\R{$\mathscr{R}$}
\def\mR{\mathscr{R}}

\def\mN{\mathds{N}}

\def\mRR{\mathds{R}}

\def\csum{\,\sharp\,}

\newcommand{\bra}[1]{\langle #1|}
\newcommand{\ket}[1]{|#1\rangle}

\def\spacing{1}
\def\bls{\renewcommand{\baselinestretch}}
\bls{\spacing}

\begin{document}
\title{~\\Unified meta-theory of information, consciousness, time\\
  and the classical-quantum universe\\~}

\author{\normalsize Martin A. Green\thanks{E-mail:
    \href{mailto:mgreen@perimeterinstitute.ca}{mgreen@perimeterinstitute.ca}
    \protect}\vspace{6pt}\\
  \normalsize \emph{Perimeter Institute for Theoretical Physics}\\
  \normalsize \emph{Waterloo, Ontario ~N2L 2Y5, Canada}}

\date{April 28, 2014}

\maketitle

\bls{1}
\begin{abstract}
  \noi As time advances in our perceived real world, existing
  information is preserved and new information is added to history.
  All the information that may ever be encoded in history must be
  \emph{about} some fundamental, unique, atemporal and pre-physical
  structure: the bare world.  Scientists invent model worlds to
  efficiently explain aspects of the real world.  This paper explores
  the features of and relationships between the bare, real, and model
  worlds.  Time\,---\,past, present and future\,---\,is naturally
  explained.  Both quantum uncertainty and state reduction are needed
  for time to progress, since unpredictable new information must be
  added to history.  Deterministic evolution preserves existing
  information.  Finite, but steadily increasing, information about the
  bare world is jointly encoded in equally uncertain spacetime
  geometry and quantum matter.  Because geometry holds no information
  independent of matter, there is no need to ``quantize'' gravity.  At
  the origin of time, information goes to zero and geometry and matter
  fade away.
\end{abstract}
\bls{\spacing}

\newpage
\small
\begin{quotation}
  \noi\begin{center}
    \emph{\parbox{5.25in}{\raggedright One path only is left for us to
        speak of, namely, that It is. In it are very many tokens that
        what is, is uncreated and indestructible, alone, complete,
        immovable and without end. Nor was it ever, nor will it be;
        for now it is, all at once, a continuous one. For what kind of
        origin for it will you look for?  In what way and from what
        source could it have drawn its increase?  I shall not let thee
        say nor think that it came from what is not; for it can
        neither be thought nor uttered that what is not is.  And, if
        it came from nothing, what need could have made it arise later
        rather than sooner?  Therefore must it either be altogether or
        be not at all.}}
  \end{center}
  \hspace*{\fill}Parmenides of Elea, On Nature, $5^\mathrm{th}$~c.~BCE\\
  \hspace*{\fill}translated by John Burnet, 1892
\end{quotation}
\begin{quotation}
  \noi\begin{center}
    \emph{Reality cannot be found except in One
      single source, because of\\the interconnection of all things
      with one another.}
  \end{center}
  \hspace*{\fill}Gottfried Leibniz, Philosophical Investigations, 1670
\end{quotation}\normalsize

\section{Introduction}\label{S:Intro}

Evolution has enabled our minds to efficiently comprehend the world as
a synthesis of elements and to build rational models of how those
elements unite to yield the whole.  From ancient times, philosophers
have viewed the world in terms of elemental parts or principles; for
example: earth, water, air and fire.  Complex, higher-level
structures, including cognitive beings like ourselves, are presumed
synthesized from these elements, explained by the elements' intrinsic
properties and mutual relationships and interactions.  Modern physics
has adopted geometrical spacetime and numerous species of quantum
matter as its primitive elements, with an action function and least
action principle to relate and govern them.  But the quantum
measurement problem \citep{1963AmJPh..31....6W, 2004RvMP...76.1267S,
  Wallace_2008, Sudarsky2011} and conceptual incompatibility of
general relativity and quantum theory, especially with respect to time
\citep{Ish4602856}, remain critical barriers to consistent explanation
of the high-level structures we directly perceive and of the universe
as a whole.

The fundamental nature of the elements is axiomatic in synthetic
worldviews.  Individually, however, the elements are useless; any
relevance of an element to the world depends on its relationships with
other elements, expressed as laws of nature.  Thus, in quantum field
theory (QFT), \emph{physical} (or \emph{dressed}) properties and
instances of the various species of quantum matter arise from
\emph{virtual} (or \emph{bare}) interactions involving all species;
the measured charges, masses and momenta of elementary particles are
the collective result of their indiscernible virtual interactions.

Conceptual barriers seem to block consensus on how the properties of
quantum matter and spacetime geometry are related, while respecting
the principles of general relativity.  We measure and perceive
distances and times as c-number quantities, but there are no generally
accepted means for a classical environment to emerge from quantum
matter or for observations of quantum matter to yield c-number
outcomes.  Explanations based on decoherent or consistent histories
\citep{Gell-Mann1993, Griffiths:976359, 2011arXiv1105.3932G},
environment induced decoherence \citep{zurek-2003-75,
  2007PhRvA..76e2110Z, 2011arXiv1103.5950F} and quantum Darwinism
\citep{2006PhRvA..73f2310B} must assume existence of the global
classical environment they were trying to avoid
\citep{2011AIPC.1327...26S,Sudarsky2011, 2011arXiv1102.2826F}.  From a
general relativity perspective, even claiming that a quantum system
has some spatial locale or is in some spin state requires an implied
but unexplained binding between position and angular momentum
operators on the quantum Hilbert space and the quasi-classical matter
through which the spacetime geometry and its inertial frames can be
known.\footnote{We use the term \emph{quasi-classical} to refer to
  properties of quantum matter that have become \emph{real}\/ by
  influencing the perceived world of conscious observers.  While the
  influences on different observers' perceptions must be mutually
  consistent, the representation of abstract information about the
  matter in terms of observable properties of the world will so
  entangle the bits that they are individually inaccessible (i.e.,
  entropy).}  The dynamical laws of quantum theory and general
relativity, considered separately, relate the ways in which given
information (or prior knowledge) is represented at different times,
without adding, removing, or otherwise changing the
information\,---\,they are both deterministic theories.  The
non-deterministic, probabilistic character usually associated with
quantum theory actually enters only in the state reduction, or
\emph{measurement}, process that links the quantum and classical
worlds: the mysterious quantum-classical transition.

Although the synthetic approach has been fruitful, there has always
been a desire to attain a fully unified view of the world, with
fewer\,---\,ideally just one\,---\,primitive elements.  Einstein
showed how abandoning the \mbox{Newtonian} elements, space and time,
in favour of principled analysis could reveal important new physics of
unified spacetime.  The ultraviolet catastrophe, implied by a unifying
application of classical statistical mechanics to electromagnetic
radiation, led to the discovery of quantum phenomena.  Soon all matter
was viewed as quantum mechanical and QFT was developed to unify
quantum theory with special relativity, forcing abandonment of
particles as immutable elements.  The standard model of particle
physics builds on a QFT framework to give a unified account of all
observed elementary particles and their interactions.

In this paper we step back from the established frameworks of the
synthetic worldview.  We reconsider the nature and relationships of
basic concepts such as information, consciousness, history and time.
Our arguments are intuitive and mostly analytic:
\begin{enumerate}
\setlength{\parskip}{3pt}
\setlength{\topsep}{0pt}
\setlength{\itemsep}{0pt}
\item I am a conscious observer, aware of my present real world which
  includes me and other conscious observers.  I am also aware of
  changes, that I associate with progress of time, and their
  cumulative contributions to the history of my world.
\item The real world $\mR_A$ that a conscious observer $A$ may
  perceive, or that may have in some way influenced $A$'s present
  state of being, has exactly the same information as the complete
  history of $A$'s present.\footnote{\emph{Information} is always
    \emph{about} something, say $X$.  Complete information about $X$
    can be identified with $X$: $I(X)\equiv X$.  If $I(Y)\subseteq
    I(X)$ then $Y$ is a representation of $X$.  If $I(Y)=I(X)$ the
    representation is faithful, otherwise it is unfaithful.}  History
  and the present real world are faithful representations of each
  other; they are just different conceptions of the same information.
\item Information that is part of present history will be preserved in
  the future even though most details of it may be inaccessible to
  conscious observers (e.g., as the unknowable actual microstate of a
  high entropy macrosystem).
\item History, and therefore $\mR_A$, can be identified with
  incomplete, but progressively accumulating, information about some
  particular thing {\B}.  Every $\mR_A$ must be an unfaithful
  representation of this same {\B}.  If time $t_1 < t_2$,
  $\mR_{A_{t_1}}$ must be an unfaithful representation of
  $\mR_{A_{t_2}}$.  (No inference should be made here regarding a
  relationship of $\mR_A$ or {\B} to conceptual models of the physical
  world involving matter and geometrical spacetime.)
\item {\B}, as conceived here, is ontologically prior to observers,
  perception, time and the real world.  Therefore {\B}, from which
  \emph{our} real world emerges, must be unique and possess more
  information than will ever be represented by our real world.  Other
  observers and real worlds could emerge from the same or other
  {\B's}, but that is irrelevant to us.
\item The advance of time can be identified with both the growth of
  history and the addition of information about {\B} to the observer's
  perceived real world.
\item Since $A$ is included in $\mR_A$, $A$ cannot infer or reliably
  predict specific information about {\B} that is not (yet)
  represented in $\mR_A$.  Therefore, for $A$, the advance of time /
  growth of history must be a random (unpredictable,
  non-deterministic) process.
\item At earlier times, $\mR_A$ will have represented progressively
  less information about {\B}, with the limit of zero information
  corresponding to the origin of time.
\end{enumerate}

\noi Our unique {\B} is equivalent to Parmenides' ``It'' or the ``One
single source'' of Leibniz.

Taking {\B} as the bare foundation, Section \ref{S:Framework}
introduces a consistent view of the world with three,
hierarchically-related levels: bare, real and model worlds.  Section
\ref{S:Models} then focuses on models, examining the implications of
this analytic worldview for quantum theory, general relativity, and
their integration and interpretation.  Recognizing real world
information as being about the bare world allows unified explanation
of time, quantum uncertainty, the low entropy state of the early
universe, and more.  Section \ref{S:Summary} summarizes our new
worldview and its implications.  A tentative proposal for the
structure of {\B} is presented in Appendix \ref{S:Topoverse}, with
arguments regarding its natural fit with quantum field theory and the
general properties of elementary particles.

\section{Comprehending the world}\label{S:Framework}

We propose that the world should be considered at three levels, with
hierarchical relationships and very different conceptual natures:

\begin{description}[topsep=2pt, parsep=2pt]
\item[W-1 ~] \parbox[t]{5.32in}{The pre-physical \emph{bare world},
    {\B}, is the unique foundation for all that might ever be
    perceived in the universe.}
\item[W-2 ~] \parbox[t]{5.32in}{Perceived \emph{real worlds}, {\R},
    arise or emerge as nested, incomplete representations of {\B}, in
    which participatory, conscious observers discover and interact
    with each other and the \emph{dressed} matter of the physical
    universe.}
\item[W-3 ~] \parbox[t]{5.32in}{Mathematical and interpretational
    models\,---\,\emph{model worlds}\,---\,are devised by conscious
    observers to efficiently explain, or account for, their past and
    future perceptions.}
\end{description}

\noindent All that is perceived as reality, including the physical
universe and all living things, evolving in time, is confined to W-2.
Physics theories, such as quantum theory, general relativity, and the
standard model of particle physics, appear in W-3.  Time is explicitly
absent in W-1, arises as an integral feature of W-2, and is
represented in \mbox{W-3} according to the needs of each model.

\subsection*{W-1: Bare world}\label{S:Foundation}

Before quantum theory, it was thought that the dynamical state of the
present fully determines the states at all future times, while also
allowing us to know the past.  Bohr taught us otherwise: only the
\emph{probabilities} of specific future outcomes can be predicted, no
matter how well the present state is known.  Nonetheless, our common
intuition is that actual perceived outcomes are real, definite
properties of the physical universe\,---\,properties whose existence
becomes a robust part of real world history.  (Quantum Darwinism
\citep{2006PhRvA..73f2310B} shows how quantum theory, when applied to
idealized \emph{subsystems}, can still be consistent with this
intuition.)  It is also intuitive that history always accumulates.
Even though the observable records of history may evolve and decay,
making most information about the past inaccessible for practical
reasons, it would contradict the meaning of history if the information
it encodes were to be changed other than through accumulation.

All the information represented by the ultimate history of our real
world, all the outcomes that will actually occur throughout time, must
be about something, some ultimate structure or foundation, {\B}.
Parmenides arguments are sufficient.  All that may ever be perceived
in our universe thus stems from {\B}.  This does not imply that all
the detailed properties of {\B} will actually, or even potentially,
contribute to history or perceptions\,---\,{\B} might hold far more
information than can ever be encoded in one history, or even many
histories.  If so, which we presume to be the case, the real world
will always be an unfaithful representation of {\B}.  Regardless, we
are free to think of {\B} as the ``One single source'' that determines
all possible outcomes, whether or not they are realized in our real
world.  Borrowing language from quantum field theory, we refer to {\B}
as the \emph{bare world}\,---\,a pre-physical world, most detailed
properties of which will never be manifested in the real world.

Not only is it pre-physical, the bare world is atemporal.  There is no
time independent of {\B}, and no time embedded within {\B}.  It is
unique and unchanging.  As Parmenides wrote, it just \emph{is}.

What should we expect {\B} to be like?  To avoid ambiguity, we assume
that {\B} can be characterized as a member of some class of objects
that may be clearly defined using the language of mathematics.  It
must be unique and definite since there is no external time whose
progress would admit changes to {\B} or transition to a different
{\B}.  Generic structural properties, that admit discrete
characterization, are required for {\B} to produce a perceived
universe with a small number of dominant matter species whose distinct
properties and governing laws seem generic across (observed) space and
time.  To be the source of all the actual particles in the universe
and their detailed interactions throughout time, {\B} must have
sufficiently numerous detailed (as opposed to generic) properties.
Finally, {\B} must be highly interconnected to give rise to the causal
relationships and entanglement that make the physical universe
function as an undivided whole.

A tentative proposal, elaborated in Appendix \ref{S:Topoverse}, is
that {\B} is a connected 4-dimensional topological manifold, {\T},
with no geometry, fields or other decoration.  The properties of
interest will be the generic structural properties of 4-manifolds and
the detailed, very complicated, specific connectivity of
{\T}\,---\,nothing more.

\subsection*{W-2: Real worlds}\label{S:Reality}

If {\B} is the foundation for all potential knowledge of the real
world, then the real world perceived by any observer $A$, at her time
$t$, must be a \emph{representation}\/, $\mR_{A_t}$, of\/ \B.  All
such representations, {\R}, will be incomplete, or unfaithful; but as
the observer's time advances they will hold / encode / represent
progressively more complete information about {\B}.

Since any conscious observer $A$ must intuitively perceive herself as
part of the real world, she and all her cognitive capability,
memories, and mental models must be included within each $\mR_{A_t}$.
So also must any other observers who may exist, or have existed,
within the world that $A$ perceives at time $t$.  $A$'s perceptions
need not, and indeed will not, reflect all properties of (or
information in) $\mR_{A_t}$.  However, all of $A$'s actual perceptions
up to and including time $t$ must be attributable to $\mR_{A_t}$,
while at any later time, \mbox{$t_1 > t$}, $A$ may have new
perceptions that cannot be attributed to $\mR_{A_t}$.

It is not necessary for every representation {\R} of {\B} to contain
observers, but observers are essential to our analysis and, being part
of reality, they have no existence independent of the representations.
We will simply not concern ourselves with representations of {\B} that
do not include conscious observers.  Accommodating the observations,
intuitions and actions of observers severely constrains the features
and relationships of the representations {\R}.

As indicated above, one should think of each {\R} as the entire real
world (which we identify with its history, regardless of what details
are accessible or decipherable) at some \emph{present} time of some
actual or hypothetical conscious observer.\footnote{As cautioned in
  Section \ref{S:Intro}, no inference should be made regarding a
  relationship of {\R} to, for example, the causal past of a point in
  geometrical spacetime, which involves a conceptual \emph{model} of
  the physical world.}  Let $\mR_{t_1}$ and $\mR_{t_2}$ correspond to
$A$'s two times \mbox{$t_1 < t_2$}.  Since history always accumulates,
$\mR_{t_1}$ must be an unfaithful representation of $\mR_{t_2}$, which
we denote by \mbox{$\mR_{t_1} \prec \mR_{t_2}$} or \mbox{$\mR_{t_2}
  \succ \mR_{t_1}$}.  Now let $B$ be a second observer such that $A$
and $B$ have, at their respective times \mbox{$t \ge t_0$} and
\mbox{$u \ge u_0$}, memories of each other.  Since $A$ and $B$ are
contemporaries, there must exist times $u_1$ and $t_1$ such
that:\\
\parbox{5.5in}{\begin{eqnarray*}
    ~~\mR_{B_u} \prec \mR_{A_{t_0}} \mathrm{~for~} u \le u_1, &
    &\mR_{B_u} \not\prec \mR_{A_{t_0}} \mathrm{~for~} u > u_1, \\
    ~~\mR_{A_t} \prec \mR_{B_{u_0}} \mathrm{~for~} t \le t_1, &
    &\mR_{A_t} \not\prec \mR_{B_{u_0}} \mathrm{~for~} t > t_1.
  \end{eqnarray*}}\eqnarraynum

If \mbox{$\mR_{A_t} \not\prec \mR_{B_u}$} and \mbox{$\mR_{B_u}
  \not\prec \mR_{A_t}$} then $A$'s real world at her time $t$ encodes
information about {\B} that is not encoded in $B$s' real world at his
time $u$, and vice versa.  In that case, $\mR_{A_t}$ and $\mR_{B_u}$
are temporally unordered.

The representations {\R} are nested and overlapping in a manner
analogous to observers' past light cones (including the interiors) in
general relativity.  But each {\R} is here understood, instead of as a
geometrical object in spacetime, as the entire real world (not just
the obviously physical aspects) at the perceived present time of an
observer.  All information about {\B} held in each {\R} has been woven
into the history and memories whose culmination is the observer's
present.  Our deep sense of time is rooted in the partial order
``$\prec$\,'' that organizes this information.

Progress of time corresponds, identically, to accumulation of history,
including new memories for the observer.  Each addition to the history
of the real world perceived by $A$ represents information about {\B}
that was not represented in $\mR_A$ at earlier times.  Although it is
eternal in {\B}, this is always truly new information in $\mR_A$, and
as such it is, in principle, unpredictable by $A$.  This
unpredictability is the underlying reason for uncertainty of quantum
measurement results.

The status and distinct characteristics of the past, present and
future are natural consequences of this worldview.  The \emph{present}
is always the complete real world {\R} perceived by a conscious
observer.  An observer's perceived present exists, in essence, as the
mental models and memories in her mind.  The \emph{past} is contained
within the present as a conceptual means to efficiently organize and
store (incomplete) information about {\B}.  The observer's conceptual
past times, $t_0$, serve to label the representations $\mR_{t_0}$
related by the partial order ``$\prec$\,'' to the present $\mR_t$, for
$t_0 < t$.  This ensures that a bit of information about {\B} that is
first captured as part of reality in $\mR_{t_0}$ is efficiently
maintained as part of the more complete representations that are
perceived as later times.  The \emph{future} is a conceptual feature
of our mental models that ensures the integrity of history will always
be maintained as we continue to discover and consistently incorporate
more information about {\B} into {\R}.

That reality should be subject to dynamical laws can be attributed to
the organizing principle\,---\,the hierarchical organization of
history\,---\,applied to efficiently encode information about {\B}.
To maintain consistency while incorporating additional information
requires the detailed representation of history to be continually
adjusted.  Preservation of prior information necessitates
deterministic classical dynamics and deterministic evolution of
quantum states (which encode prior knowledge).  Our predictions of the
future are perhaps best thought of as conditions the future must
satisfy to consistently preserve present history.  Probabilistic
quantum predictions must anticipate, and be compatible with, all
possible consistent outcomes and the \emph{hypothetical} new
information about {\B} they represent.  Actual future outcomes must
correspond to \emph{actual} information about {\B}.  While all
hypothetical outcomes must be compatible with the generic properties
of {\B}, most of those hypothetical outcomes will be incompatible with
the actual details of {\B}; but there is no way of knowing, in
advance, which ones to exclude.

One might view {\R} as conceptually similar to a (crystallizing or
evolving) block universe \citep{2010IJTP...49..988E}, but there are
major differences worth noting.  Most importantly, $\mR_t \prec \mR$
does not imply that {\R} is an extension (or superset) of $\mR_t$ but,
instead, that all information about {\B} encoded in $\mR_t$ is also
encoded in {\R}.  The detailed form of the encoding will generally be
different, and detailed disentangling of old from new information in
{\R} will generally not be feasible.  Whereas block universe models
are focused on matter in spacetime, such physics-specific structure is
not assumed for {\R} but appears instead in the model worlds of W-3.

An obvious question, now, is: what caused conscious beings to arise in
the representations {\R} of {\B}\,?  The simple answer is: because
they could; because {\B} has sufficient structure and complexity to
have this wonderfully rich, hierarchically linked family of
representations within which consciousness could emerge.  Most
representations of {\B} will not likely be so fruitful.  But we and
our perceived world are all the evidence we need to conclude that
consciousness arises in at least one family of representations.  There
might be many other real worlds, even many other {\B}'s, but they will
have no bearing on us.

Questions regarding free will are also relevant here.  Do conscious
beings have free will\,---\,the ability to consciously influence their
future?  Or is their future fully determined by {\B} and free will
just an illusion?  I claim that free will is both necessary and
natural.  As new information is acquired and consistently integrated
with prior information, each new present, $\mR_t$, corresponds to
$A$'s progressively more complete representation of {\B}.  But the
order in which new information about {\B} will become part of $A$'s
real world is not determined by {\B}, nor by $A$'s present.  Free will
is a manifestation of the ability of $A$, and other conscious beings
who are part of $A$'s present, to influence the \emph{order} in which
previously unknown information about {\B} is consistently incorporated
into their reality, including their mental models thereof.  This is
roughly analogous to a conscious observer's progressive development,
by shifting their viewpoint and the illumination, of a mental model of
an object (\R) from the information encoded in a holographic plate
(\B).  The big differences, here, are: (i) the conscious observer is
part of {\R}; and (ii) there is so much information in {\B} that only
a tiny portion of it will ever be represented in {\R}, allowing
different paths (viewpoint and illumination) to yield materially
different outcomes.

Acceptance of free will leads to a further consistency challenge: a
willful action of $A$ at her time $t$ must make compatible
contributions to history for all observers $B$, at their times $u$
such that \mbox{$\mR_{A_t} \prec \mR_{B_u}$}.  Otherwise, history
would become inconsistent.  This leads to a somewhat idealist view, in
which consciousness has a global, collective aspect rather than
arising completely independently in each observer.  With this view, we
must recognize as illusion the intuition that we, as individuals, have
completely separate psychological existences.  I am not suggesting
that there is no independence, only that its scope is limited.  Our
pervasive, underlying entanglement with each other and with our
perceived physical environment\,---\,the globally distributed encoding
of information about {\B}\,---\,ensures that our individual willful
actions have global effect.

\subsection*{W-3: Model worlds}\label{S:Explanation}

The perceived real world, conscious observers, history and time emerge
together as an incomplete representation of {\B}.  Everything real is
fully \emph{dressed}\,---\,as opposed to \emph{bare} or
\emph{virtual}.  The robust, causally consistent history of the
perceived universe is reflected in the content and organization of
observers' memories.  In imagining the future, an observer preserves
history while judging, and to a limited extent influencing, how
history might be extended by unpredictable new information about {\B}.

The reality described above is intuitive, yet mysterious.  As
intelligent observers, we invent mathematical and interpretational
models to rationally, efficiently, and reliably simulate selected
features of reality.  These models constitute W-3.  While they
simulate (features of) the perceived reality of W-2, valid models must
also be consistent with and meaningfully, although perhaps not
obviously, reflect the properties of {\B}.  Reflecting the generic and
specific properties of {\B} and {\R}, models may take the form,
respectively, of: (i) mathematical objects given some interpretation
and subject to constraining \emph{laws} (e.g., general relativity,
quantum theory, and the standard model of particle physics), or (ii)
instances of specific structure (i.e. particular solutions) within the
context of a generic model (e.g., the concordance model of cosmology
($\Lambda$CDM)).

The laws of physics are understood here as constraints that apply
within observers' models, necessitated by the generic properties of
{\B} and the generic perceived features of the representations {\R}.
Since they apply only to models, the laws of physics do not, in any
way, constrain {\B} or {\R}.  But if a model is based in part on
conjectures regarding the generic properties of {\B}, and then found
to reliably simulate certain aspects of the real world, this will give
credence to the conjectures and justify their further use to
\emph{explain} the real world.

Concepts of time, 3-dimensional space, associated geometry, and matter
whose spatial distribution evolves in time are entwined within our
mental models of physical reality.  These concepts are naturally
reflected in the mathematical models of \mbox{Newton}, Maxwell, and
Einstein, which reliably simulate aspects of reality within their
respective domains of applicability.  These deterministic models do
not, however, simulate the addition of new information about {\B} into
\R.\footnote{The measured isotropy of the CMB implies that the entire
  observable universe is causally connected\,---\,there is no part of
  our universe, yet to be observed, whose corresponding prior
  information is not already represented in our history.  Models of
  cosmic inflation were developed to explain how this could be.}
Quantum models of matter provide a mechanism\,---\,state
reduction\,---\,to admit such new, unpredictable information.

Models are discussed in more detail in Section \ref{S:Models}.  The
key message here is that models do not govern reality.  They simulate,
and thus represent, aspects of reality; but the representations are
always unfaithful.  This implies that all models, and hence all
physical laws, have limited domains of applicability, which we need to
discover and accept.

\section{Models\,---\,Interpretation and Implications}\label{S:Models}

Models are the product of science\,---\,created by intelligent beings
who exist only as emergent features of the real world.  Scientists
develop mathematical and interpretational models to efficiently
simulate certain aspects of real or hypothetical worlds.  Theoretical
constructs such as geometrical spacetime and quantum matter prove
useful as elements of models because they can be mathematically
precise and they can be interpreted in a manner that establishes
reasonable correspondence with certain perceived properties of the
real world.  A theory defines the generic properties of a class of
model worlds, whereas a particular solution corresponds to some
specific model world.

Current physics theories typically represent information about the
\emph{generic} structure of {\B} in terms of: the identities of
fundamental matter fields; their spacetime and internal symmetry
properties; and the principles and laws that govern their interaction
and evolution.  Applications of a model often rely on prototype
solutions, such as black holes or idealized quantum subsystems
(elementary particles, nuclei, atoms, etc.).  These proto\-types are
derived as abstract solutions of the model's underlying theory,
without reference to specific real world observations; they may be
consistent with the generic structure of {\B} but do not hold any
specific information about {\B}.

Modeling of \emph{specific} properties of the real world requires
particular solutions of generic models, plus an interpretation that
establishes reliable correspondence between the solution and those
real world properties.  Such solutions can then be interpreted as
encodings of the subset of information about {\B} that corresponds to
the pertinent real world properties.  The concordance model of
cosmology serves as a good example.  The parameter values and
interpretation that yield the best correspondence between the model
and observations encode information about {\B} that is represented in
$\mR_t$ as specific, coarse-grained properties of the real
world\,---\,properties that are preserved in history from early times
and for all known observers.

Like cartoons, models provide an idealized view of certain selected
aspects of the real world.  Every model thus has a limited domain of
sensible correspondence with the real world, even though the limits
may not be obvious.  Stretching a model beyond its domain, or
simultaneous application of two incongruous models, may produce
confusing or unreasonable results\,---\,hence the difficulty of
extending quantum theory and general relativity into each other's
natural domain.

\subsection{Time}\label{SS:Time}

The new conception of time in W-2 makes it necessary to revisit our
interpretation of the correspondence between the times of classical
and quantum physics models and an observer's past, present and future
in the real world.  In particular, our interpretation must address how
models simulate preservation of information that is already included
in {\R} and how they simulate addition of new information to {\R} as
time progresses.

Deterministic theories provide the means to simulate the
representation of given, fixed information about {\B} in terms of
robust properties of the real world\,---\,properties that are
redundantly expressed as discussed in \citep{2006PhRvA..73f2310B}.  A
quantum state $\psi$ always represents precisely the same (prior)
information at all hypothetical times\,---\,past, present and future.
To sensibly associate quantum models with specific properties (e.g.,
idealized quantum subsystems) of the real world we implicitly assume
an association of quasi-classical, time-space-matter descriptors of
the real world with corresponding self-adjoint operators on Hilbert
space.  (In the Heisenberg representation, unitary evolution is
imposed through the time-dependence of the assumed associations
between the quantum operators and the quasi-classical world,
independent of $\psi$.)  Existence of the quasi-classical world is a
corequisite for such associations to be sensibly made.  The
associations define a correspondence between information represented
in (quasi-classical) history and information represented in quantum
states and operators.  No actual quantum state can be prepared or
known other than through the associations discussed above and the
additions to history that reflect the state's preparation and
measurement.  Moreover, all prior information is quasi-classical,
having been collectively and non-locally entangled in the (mostly
inaccessible) details of history.

Because they can neither add nor remove information, but only change
the details of its representation at different times, deterministic
theories cannot distinguish the present or identify a preferred
direction of time.  They are suited to modeling the evolving
time-space-matter representation of already-represented information
about {\B} in a manner that decouples it from the acquisition of new
information and the growth of history.  Such decoupling is required in
all deterministic models, but it is an idealization with limited
validity since it denies the growth of history.  True progress of time
always involves enriching {\R}, and thus history, with unpredictable
new information about {\B}.  Increasing information (and growing
history) defines the preferred direction of time.  New information
will be manifested in the real world with an encoding that necessarily
entangles new and prior information, and that no deterministic model
can simulate.  Only through a non-deterministic process can a model
simulate the required introduction of new information.

Quantum state reduction is precisely the needed non-deterministic
process.  When we focus on a particular subject quantum subsystem, our
working assumption is that time advances independently.  But for this
to be valid, some other subsystem(s) must undergo state
reduction\,---\,clocks must tick\,---\,thereby progressively enriching
history and increasing the algorithmic information content (or
Kolmogorov complexity \citep{Ming93}) of the quasi-classical
quantities in terms of which the reduced subsystems and operator
associations are defined.  Each such state reduction will add at least
one \emph{bit} of information about {\B}.  Each new bit will
correspond to one Planck unit of action ($\hbar$) that irreversibly
changes the state of the quasi-classical universe.  Since the Hilbert
space of quantum subsystems must be defined and interpreted through
reference to quasi-classical descriptors of the perceived world, the
increasing complexity of the quasi-classical universe will also
steadily increase the size of the total Hilbert space, of which a
(now) time-dependent subspace corresponds to our subject subsystem.

By construction, any subject quantum subsystem represents certain
limited prior information encoded in history at some reference present
time.  Attempting to assign likelihoods to different (past) histories
consistent with this limited subsystem would be counterproductive,
since our actual, full history, including the inaccessible details of
its microstate, is what defines the complete present.  (Our inability
to resolve details or disentangle the present to unambiguously
reconstruct the history of earlier times does not make present history
any less definite.)  Applying the Born rule (or, equivalently, Bayes
theorem \citep{2002quant.ph..5039F}) to determine the likelihood of
potential future outcomes related to our subject subsystem is
sensible, provided there is sufficient isolation so that the validity
of the subsystem's unitary evolution will not be significantly
impacted by entanglement with, and unpredictable state reduction of,
other quantum subsystems.  New information represented in our actual
future, as it comes, will be reflected in the consistent encoding of
its history; the likelihood of each potential history is identically
the likelihood of the potential future to which it corresponds.

A notable prior attempt to understand physics in terms of steadily
increasing information is Von Weizs\"{a}cker's \emph{ur\/-theory}
\citep{VonWeizsacker:1250894,1996quant.ph.11048L,2003tqi..conf..375L}.
Following the development of quantum information theory it was
realized that \emph{urs} are the same as \emph{qubits}.  The time
operator in ur\/-theory is just the \emph{ur\/-number} operator, which
gives the number of bits.  In our analysis, state reduction transforms
qubits to Planck unit steps in action, whose random walk determines,
with suitable coarse-graining, the quasi-classical action of the
perceived world.  The relation of the amount of accumulated
information\,---the number of bits gained through elementary state
reductions\,---\,to metrical time is monotonic, just as in ur-theory.

Interpretations of orthodox quantum theory (e.g., environment induced
decoherence, quantum Darwinism, consistent histories) attempt to
explain the apparently classical universe while avoiding state
reduction as a distinct process.  Collapse theories (see
\citep{sep-qm-collapse} and references therein) introduce state
reduction as a stochastic nonlinear dynamical process that may involve
gravity \citep{1989PhRvA..40.1165D,penrose1989emperor,Penrose00}.
Penrose considered a possible entropic relationship between such state
reduction and its influence on spacetime geometry \citep{Penrose86}.
In almost all these approaches, time is presumed to exist independent
of the state reduction events; and none recognizes that state
reduction events must add information that progressively accumulates.
An exception is a simple model proposed by Pearle
\citep{2013FoPh...43..747P} in which, starting from a ``nothingness''
state, collapse events generate discrete time and space.  Our position
is that state reduction is the essential process through which time
progresses and information is added to history.  Since the information
is nonlocally encoded in the inaccessible details of the microstate of
the world, almost all state reductions are noticed only as the passage
of time\,---\,something changed, enriching history.

\subsection{Uncertain reality\,---\,matter and
  geometry}\label{SS:Semiclassical}

The dominant view of physicists today is that the universe is
fundamentally quantum.  Arguments based on environment-induced
decoherence and coarse-grained histories show how an orthodox quantum
model can lead to observations of matter that are apparently
quasi-classical.  Essentially, the environment gains sufficient
information from its entanglement with a subsystem to fully describe
the subsystem's decohered, quasi-classical aspects.  This information
is claimed to be ``somewhere in the universe''
\citep{Gell-Mann1993,Halliwell1999}.  Of course, that claim relies on
implicit correspondence between the model and the real world.  Quantum
Darwinism shows that, given such correspondence, this information will
be accessible throughout the universe (subject to causal constraints),
thereby ensuring that different observers will agree on
quasi-classical outcomes\,---\,their perceived real worlds will be
mutually consistent \citep{2006PhRvA..73f2310B}.

Most environment and apparatus degrees of freedom that are
\emph{traced over} to focus attention on a subject quantum subsystem
are inaccessible to classical observers; their measurement is not
feasible.  The \emph{amount} of inaccessible information encoded in
the actual, but unmeasurable, microstate of the world associated with
these degrees of freedom is referred to as entropy.  Remaining
subsystem degrees of freedom may be measured, with residual
uncertainty, by entangling the subsystem with the apparatus and
environment.  The timing of such measurements depends on the
apparatus; some subsystem degrees of freedom may be measured sooner
than others.  For example: independent measurements of the two
particles of an EPR pair may localize their uncertain energy-momentum
densities within limited spatial regions that have spacelike
separation.  Until a later spin measurement is performed on one of
these particles, their carefully isolated collective spin state (which
is known by virtue of the preparation process) remains bound to the
pair but itself has no locale.  Following the spin measurement, the
reduced state represents both the prior information, encoded in the
collective spin state, and new information encoded in the measured
spin.  Violation of the Bell inequalities is necessary because a spin
measurement on the second particle must also preserve the same prior
information.

General relativity establishes a mutual relationship between classical
matter and spacetime geometry, with the idealization that both are
characterized by smooth, real-valued tensor fields on a 4-manifold
{\M}:
\beq%
G_{\mu\nu}(x) = 8\pi G_{\!N} T_{\mu\nu}(x)\,,~x\in\mM\,,\label{Eq:Einstein}%
\eeq
where $G_{\mu\nu} = R_{\mu\nu} - \frac{1}{2}g_{\mu\nu} R$ is the
Einstein tensor corresponding to the metric $g_{\mu\nu}$, and
$G_{\!N}$ is Newton's gravitational constant.  (We use units in which
$\hbar = c = k_B = 1$.)  A cosmological constant term is not excluded,
but left implicit in the stress-energy tensor, $T_{\mu\nu}$.
Observations to date are consistent with {\M} being simply connected.
Equation \eqref{Eq:Einstein} demands consistency of dynamical matter
and spacetime geometry on any Cauchy surface and, when combined with
deterministic equations of motion for the matter, constrains the
evolution of geometry to encode precisely the same information at
other times.

Even if there is no matter ($T_{\mu\nu}=0$), use of real-valued fields
implies that the information encoded in the geometry of a generic
solution of the Einstein equations will be uncountably infinite.  We
have argued, however, that the amount of information represented by
the real world equals the finite number of elementary state reductions
since the origin of time.  Generic spacetime geometries thus encode
far too much information to be good models of the real world.  If we
want a model that reflects the finite, growing information content of
the real world we must reconsider the geometrical spacetime
interpretation.

Recall that $A$'s real world $\mR_{A_t}$, for finite $t$, encodes a
finite amount of information about {\B}.  The algorithmic information
content of the corresponding model-space, expressed as bits, should
then be the number of elementary state reduction events in $A$'s
entire causal past.  Equation \eqref{Eq:Einstein} tells us that the
information encoded in $G_{\mu\nu}$ (or $R_{\mu\nu}$) is identical to
the information about matter encoded in $T_{\mu\nu}$.  Moreover, our
``knowledge'' of spacetime geometry is entirely inferred from
observations of matter, so it is reasonable to expect that the
information represented by geometry corresponds identically to
\emph{all} the information represented by matter.  This means that
Weyl curvature components, which when combined with Ricci curvature
fully determine the metric geometry, must be non-zero only to the
extent required by causality and the Bianchi identities, $\nabla_\nu
G_\mu^{~\nu}= 0$\,.  Any Weyl curvature that is not required for
causal consistency with the stress-energy tensor of matter sources
would encode information that has no relation to {\B}.  Solutions of
the Einstein equations with source-free Weyl curvature, or with
sources that persist to the origin of time, should be considered
non-physical.  This includes primordial gravitational waves and
primordial black holes.

To properly model the nesting of real worlds discussed in Section
\ref{S:Reality}, the field $T_{\mu\nu}(x)$, for all points $x$ on and
within the past lightcone of $A$'s perceived spacetime location
$x_{\!A}(t)$, must encode (a subset of) the information represented in
$\mR_{A_t}$.  Since a given (nearly) isolated quantum subsystem must
have little influence on geometry relative to the the influence of the
rest of the universe, we adopt the decoherence approach and expand the
stress-energy tensor of equation \eqref{Eq:Einstein} as:
\beq%
T_{\mu\nu}(x) = {T_\mathrm{e}\,}_{\mu\nu}(x) +
\langle\hat{T}_{\mu\nu}(x)\rangle_\mathrm{s}\,,\label{Eq:semiclassical}%
\eeq
where ${T_\mathrm{e}\,}_{\mu\nu}(x)$ is the perceived quasi-classical
stress-energy tensor of the environment (the rest of the universe),
excluding the quantum subsystem of interest, and the expectation
value:
\beq%
\langle\hat{T}_{\mu\nu}(x)\rangle_\mathrm{s} \doteq
\bra{\psi_\mathrm{s}} \hat{T}_{\mu\nu}(x)\ket{\psi_\mathrm{s}}%
\eeq
corresponds to $A$'s prior knowledge of the subsystem's stress-energy
operator $\hat{T}_{\mu\nu}(x)$ and state $\psi_\mathrm{s}$.

The environment-subsystem split indicated in \eqref{Eq:semiclassical}
can be freely chosen to treat as quantum as much of the universe as
can be effectively isolated (with minimal decoherence of its quantum
properties by the environment).  No matter where the split is chosen,
the combined environment and subsystem will always represent exactly
the same prior knowledge and the same total stress-energy and
geometry.  In practice, however, for any subsystem that can be
effectively isolated the geometry will be overwhelmingly determined by
${T_\mathrm{e}\,}_{\mu\nu}$.  For all practical purposes, this
dominance of the environment allows the needed association to be made
between $x$ and $\partial/{\partial x}$ in $\hat{T}_{\mu\nu}(x)$ and
spacetime coordinates and inertial frames defined by reference to
quasi-classical matter of the environment.

In our new worldview, emergent geometry derives all its information
from non-unitary state reduction of quantum matter.  There are no
independent geometrical degrees of freedom.  State reduction causes
$T_{\mu\nu}$ to evolve in a stochastic manner, but the fluctuations
are imperceptible because of the vast amount of information already
encoded in the observable universe.  Existing information (prior
knowledge) encoded in ${T_\mathrm{e}\,}_{\mu\nu}(x)$\, is always
preserved, even though it does not fully determine the future.  With
only finite information, the representation of stress-energy and
geometry in terms of real valued fields on a 4-manifold must be
intrinsically uncertain, which is consistent with the quantum
uncertainty of $\langle\hat{T}_{\mu\nu}(x)\rangle$ that has
been studied in the context of stochastic semi-classical gravity
\citep{lrr-2008-3}.

A Feynman sum-over-histories approach can be used to predict the
likelihoods of future outcomes, but our new interpretation imposes
some new restrictions:
\begin{itemize}
\setlength{\parskip}{3pt}
\setlength{\topsep}{0pt}
\setlength{\itemsep}{0pt}
\item Allowed histories must have monotonically increasing information
  content.  This includes both accessible and inaccessible (entropy)
  information.
\item Each consistent history must be determined solely by its matter
  degrees of freedom and associated state reduction events, with the
  geometry implicitly matching the matter.  Equation
  \eqref{Eq:Einstein} allows determination of mutually consistent
  $T_{\mu\nu}$ and metric, $g_{\mu\nu}$, for each potential history.
\end{itemize}
For every allowable history, c-number descriptions of both matter and
geometry are uncertain, due to the finite information content, but the
uncertainties are correlated; moreover the number of Planck unit steps
in the action between initial and final times (i.e., between the
corresponding past light cones) should precisely equal the information
added to history.

An exact correspondence between uncertainties of matter and geometry
should carry over to perturbative calculations in quantum field
theory.  In particular, the compatibility between matter and geometry
imposed by equation \eqref{Eq:Einstein} should ensure that classical
uncertainty of spacetime curvature, due to its finite information
content, effectively cuts off the momenta of loop integrals to
naturally eliminate ultraviolet divergences.  This same effect,
together with the exclusion of source-free Weyl curvature modes,
should naturally limit the Casimir energy.  Finiteness of information
in the causal past of any spacetime point will also impose a natural
infrared cutoff.

Traditionally, semi-classical coupling of geometry to quantum matter
has assumed a c-number stress-energy tensor given by $T_{\mu\nu}(x) =
\bra{\psi}\hat{T}_{\mu\nu}(x)\ket{\psi}$\,, where $\psi$ is the pure,
complete Heisenberg state of the world.  But that formulation suffers
from the measurement problem because it fails to explain the $x$
dependence of the stress-energy \emph{operator}.  In the decoherence
approach advocated here, the quasi-classical domain, now with finite
information, is taken to represent the present real world.  The
quantum domain represents all potential futures allowed by the generic
structure of {\B}.

\subsection{Information, entropy and entanglement}\label{SS:Info}

Modern studies of quantum information (quantum computing, quantum
cryptography, etc.) rely on the assumption that quantum theory is the
foundation of a fundamentally correct model of physical reality.
Linkage, within that model, between quantum systems and spacetime
geometry remains unexplained.  The obstacle may be that quantum
information theory treats quantum information as properties of real
physical systems, within independently given space and time.  Our
perspective is that real physical systems, space, and time are
secondary; the bare world, {\B}, is primary.  Dressed physical systems
and corresponding spacetime geometry encode incomplete information
about {\B} in a manner that compels progressively more faithful
representation of {\B} and corresponding advance of perceived time.
Time, space and matter provide the mode of representation\,---\,they
are subservient to the information.

New information enters {\R} as an apparently random event.  This adds
to the prior information, causing a projective reduction in the space
of potential futures.  In a time-space-matter model, the mode in which
the new information is encoded expands from the event at the speed of
light.  Therefore, for this information to be preserved over time its
encoding, in terms of mutually consistent geometry and matter fields,
must be maximally non-local.

Encodings of information from different events very quickly become
highly entangled in the time-space-matter model.  With their
non-linear coupling through equation \eqref{Eq:Einstein}, the
contributions of a vast number of modes give spacetime geometry its
seemingly definite, though still uncertain, form.  Almost all matter
degrees of freedom are inaccessible to direct observation, and are
thus identified as entropy; but their specific microstate determines
the details of spacetime geometry.  Entropy associated with spacetime
geometry, such as the Bekenstein-Hawking entropy \citep{Hawking1976},
should then be identified with the global entanglement entropy of the
corresponding matter.  See \citep{Kay1998a} for an alternative
argument leading to the same conclusion.

As an example, consider the gravitational collapse of a star.  The
world of a distant observer, $A$, just after witnessing the rapid
collapse to a ``dark star'' will encode only slightly more information
than it did just before the collapse.  The corresponding spacetime
geometry of $A$'s past will be comparably uncertain before and after;
the total entropy of $A$'s world will thus also grow just a little.
Over a very long time, from $A$'s perspective, progressive state
reduction of the matter fields throughout the universe will increase
the common information content of the matter and geometry.  As the
geometry near the dark star approaches that of an idealized
Schwarzschild black hole the information and entropy attributable to
it will grow toward the limiting value, $S_\mathrm{BH} = 4\pi G_{\!N}
M^2$.  Since we have argued that there are no primordial black holes,
the information content of astrophysical black holes, such as
$\mathrm{Sgr~A^*}$, must come primarily from the baryonic matter that
collapsed to form them.\footnote{Since information is encoded in the
  global entanglement of all matter and the corresponding global
  geometry, it is formally incorrect to say that information (or
  entropy) are \emph{contained} in a star or other matter.
  Nonetheless, we can attribute the information to the star by
  imagining that removing the star and the past influences of all its
  matter and radiation would yield another universe with this much
  less information.} The corresponding entropy will be more than 20
orders of magnitude less than $S_\mathrm{BH}$ for a similar mass
\citep{Kephart2003,Frampton2008}.  This contrasts with the na\"{i}ve
notion that the entropy suddenly increases to $S_\mathrm{BH}$ during
the collapse process.  Of course, the difference is due to the horizon
and interior of a black hole being forever in the unknowable future of
all distant observers.  (Even an observer who falls into a black hole
will never see most of the event horizon or detect a burst of entropy
at the horizon.)

Given information is accessible to multiple observers only if its
encoding takes the form of quasi-classical records with associated
pointer observables.  Of necessity, many different perspectives of the
same detailed configuration will yield compatible quasi-classical
observations by different observers and at different times.  Highly
redundant encoding of pointer observables \citep{2006PhRvA..73f2310B}
makes this possible, at the cost of much more information being
inaccessible\,---\,i.e., entropy.  Advancing time increases both
accessible information and entropy.  Projecting backward, one must
conclude that the universe at earlier times had progressively less
information and entropy, with both going to zero at the origin of
time.

Studying the mutual entanglement of the modes that encode information
associated with a quantum subsystem becomes practical only when the
entanglement of those modes with the environment (including any
apparatus) is relatively insignificant.\footnote{Complete
  disentanglement from the environment is not possible.  Indeed, some
  continued coupling to the environment is essential in order to allow
  preparation and observation of the subsystem.}  Given such isolation
from the environment, spatial location is excluded from consideration
as a property of the quantum subsystem.  Entanglement of the subsystem
modes is then also freed from the nonlinearity imposed by coupling to
spacetime geometry.  The internally-relevant subsystem properties
represent prior knowledge of the observer about the \emph{nature} of
the subsystem, such as the number and type(s) of particles and their
total momentum and angular momentum.  This knowledge is obtained
entirely through the quasi-classical processes used to prepare or
otherwise identify the subsystem.

Projected future states $\psi_s(t)$, obtained by unitary evolution of
the prepared state \mbox{$\psi_s=\psi_s(t_0)$}, encode exactly the
same prior information, with the implicit assumption that the
environment evolves in a similar unitary manner.  But we contend that
advancing real-world time is always associated with new information
and corresponding non-unitary changes to the environment.  Without
those changes, time would not advance (as in the problem of time
\citep{isham1993canonical}).  Because no subsystem can be completely
isolated, new information about {\B} will eventually (depending on the
degree of isolation) be encoded as an event that involves significant
entanglement between the subsystem and its environment.  The result is
state reduction of $\psi_s$, non-linear encoding of the new
information in the quasi-classical environment, and new prior
knowledge to allow identification of a modified subsystem $\psi_{s'}$.
Although the new knowledge, and the reduced state $\psi_{s'}$, will be
consistent with the possible outcomes based on $\psi_s(t)$, the latter
has no influence on the actual outcome.

Since, conceptually, the subsystem can be expanded to treat as much as
needed in a quantum manner, our interpretation should not lead to
conflict with any current experimental tests of quantum theory.
However, our claim of finite information content and our epistemic
interpretation of quantum states may have significant implications for
quantum computing and other aspects of quantum information technology.

\subsection{Cosmology}\label{SS:cosmo}

Inflationary cosmology models, commonly involving scalar ``inflaton''
fields and designer potentials, were developed to explain how the
present universe could have become so flat, homogeneous and isotropic,
without extraordinary fine tuning of initial conditions.  Although
nothing more than a sketch can be provided here, it is expected that
our new worldview will point to a more natural replacement model for
inflation that produces the required conditions, at the end of the
``formative'' period, for subsequent astrophysical evolution and
observations.

We have argued that our real world, which we model as spacetime and
the matter within it, is nothing but a representation of information
about {\B}.  It does not objectively exist.  It started with \emph{no
  initial data}\,---\, zero total information, including zero entropy.
We should thus expect a cosmological model of the early universe to
have geometry and matter that emerge from nothing (physical or real).
With no \emph{a priori} physical laws (independent of {\B}) the mutual
relationship of geometry and matter suitable for representing {\B}
will need to be determined in a generalized Machian manner.  All
physical ``constants'', such as $G_{\!N}$, the masses of elementary
particles, and the strengths of interactions, should obtain values
based on the generic properties of {\B} and the specific information
encoded in {\R}.

Time and the amount of information about {\B} represented by $\mR_t$
increase in unison; so, as was done by Von Weizs\"{a}cker, the
(non-metrical) parameter $t$ can be chosen equal to the number of bits
of information.  With that choice, $\mR_1$ must be the origin of the
real world.  But with just one bit, there would be no quasi-classical
spacetime geometry or matter; such concepts require sufficient
information to support the environment-subsystem decomposition
discussed above.  To be viable, the environment must incur only minor
perturbation while adapting to the state reduction that encodes each
new bit.

The mutual uncertainty of geometry and matter, discussed in Section
(\ref{SS:Semiclassical}), must have been extreme at very early times.
An initial formative period was thus essential, at the end of which
the information represented by $\mR_t$ was sufficient for: (a)
environment-subsystem decomposition to be viable; (b) subsystems to be
modeled using quantum theory; and (c) quasi-classical spacetime and
matter to be related by equations \eqref{Eq:Einstein} and
\eqref{Eq:semiclassical}.  Of course, there were no conscious
observers during the formative period, so any characterization of it
is based solely on the requirement of consistency with presently
accessible historical records.  Looking back in time from the present,
spacetime geometry does not reach a singular origin; instead, geometry
becomes so intrinsically uncertain that its utility fades to nil.

Since the information in $\mR_t$ vanishes as $t\to 0$, the notion of
fine tuning never arises.  We cannot say the initial geometry was
flat, homogeneous, isotropic, or otherwise; it was simply nondescript.
Addition of information to $\mR$ is manifested in physical models as
state reduction and progress of time.  Since state reduction events
are stochastic, they should appear as random perturbations of the
emerging, but still highly uncertain, geometric environment.
(Individual perturbations will be masked by the geometric uncertainty,
thus never discernable.)  Most information will contribute to entropy,
with coarse-grain properties yielding the observables associated with
a quasi-classical ``environment''.  Each emergent mode requires a
larger scale environment (matter and geometry) within which it can be
meaningfully defined, so emergence must start at a global scale and
proceed iteratively to smaller scales, eventually resulting in an
initially scale-invariant spectrum of uncorrelated modes.  Ultimately,
the non-linear coupling of quasi-classical modes will cause growth of
density perturbations, indicating substantial completion of formation
of the geometrical environment and the matter it contains.

In parallel with geometry, matter will emerge progressively.  Species,
masses, coupling constants, modes within species, and collective modes
will acquire \emph{dressed} physical status according to the generic
and specific structure of {\B}.  The non-zero stress-energy density of
quasi-classical degrees of freedom, which makes them observable, is
exactly reflected (in value and uncertainty, through equation
\eqref{Eq:Einstein}) in the spacetime geometry.

The formative period produced initial populations of many identical
particles of each significant species.  Given these populations,
further evolution encodes new information about {\B} in terms of more
detailed interactions and distribution of the given particle
fields\,---\,the formation of nuclei, atoms, molecules, dust, stars,
galaxies, etc.  Since information and entropy can only ever increase,
the universe will never return to its initial state.  Should {\B} have
finite information then the ability to improve its time-space-matter
representation may eventually be exhausted, resulting in an effective
end of time.

\subsection{Unification}\label{S:Unity}

For over fifty years, physicists have presumed the quantum world is
fundamental and have attempted to derive the classical world from it.
The position advocated here is that neither quantum nor classical has
priority; a unified model world must have complementary, mutually
supportive classical and quantum aspects.  Their common foundation is
the bare world {\B}.  The real world that emerges is quantum
\emph{and} classical.

The formal mathematical frameworks of both quantum theory (including
quantum field theory) and general relativity are idealizations, each
of which ignores the other.  The remarkable success of the idealized,
independent theories cannot be disputed.  However, the puzzles,
paradoxes and infinities associated with them may well arise from
failure to respect their complementarity.

Theoretical predictions derived from the mathematical formalisms of
idealized theories must be considered with caution.  The enormous
entropy of Schwarzschild black holes may have no relevance to any
conceivable observations or testable predictions.  The dramatic
speed-up of quantum computers may disappear when the limits of finite
information content and the reality of state reduction are imposed.

Our new worldview, with {\B} as the foundation, reveals the origin and
complementary nature of classical and quantum.  It explains why the
universe started with zero entropy, and why the coupling of geometry
and matter must be semi-classical.  Semi-classical coupling forces
departure from the linearity of quantum theory, demands examination of
real physical states rather than abstract states, and requires real
state reduction in order to enrich history.  Deciphering the generic
structure of {\B} (with {\T} as a tentative proposal\,---\,see
Appendix \ref{S:Topoverse}) may reveal why observed matter can be
modeled by the fields and interactions of the standard model of
particle physics.  It may even reveal the nature of dark matter and
dark energy.

\section{Summary}\label{S:Summary}

Intuitive arguments regarding information, conscious perception, and
history have led us to consider the world at three-levels:
W-1\,---\,the bare world {\B} that is the unique atemporal,
pregeometric foundation of everything that may ever be perceived;
W-2\,---\,real worlds {\R} in which we perceive ourselves and
everything that seems real at each present time; and W-3\,---\,model
worlds invented by scientists to efficiently simulate aspects and
relations of the real worlds.

Progressive inclusion of information about {\B} into its emerging,
hierarchically nested representations $\mR$ generates consistent
history and gives rise to the concept of time.  Dynamical evolution of
the time-space-matter encoding of history must accommodate new
information while preserving all prior information.  The present
always corresponds to all of accumulated history, with the past
implicit in its encoding.  Possible futures must preserve information
encoded in the present, while adding unpredictable new information.
Acquisition of new information necessitates both quantum uncertainty
and state reduction.

The bare world must have sufficiently complicated generic and specific
structures to account for both the generic properties of matter and
the large number of particles and interactions in the observed
universe.  To consistently preserve history, the quasi-classical,
geometric structure of the perceived real world must encode most
information about {\B} in a form so globally entangled that the bits
become individually inaccessible (entropy).  While most entanglement
remains implicit, because observation of separate degrees of freedom
is not viable, entanglement can become explicit when studying quantum
subsystems whose few degrees of freedom are entangled more strongly
with each other than with the environment.

As models of {\R}, general relativity and quantum theory are both
idealizations whose limits must be understood in order to avoid
nonsensical conclusions.  Their integration requires adaptation of
both.  Adopting a decoherence approach to semi-classical coupling
allows both matter and geometry to have exactly corresponding
uncertainty as they jointly represent finite information about {\B}.
Geometry can have no information, or degrees of freedom, independent
of matter.  Because the information is finite: geometry cannot be
resolved at very early times and at very small distances; path
integrals become finite sums; and calculations should give finite
results in both ultra-violet and infra-red limits.

The fields, symmetries and interactions of observed elementary
particles must have their origin in {\B}.  To illustrate with a
specific example, it is conjectured, in Appendix \ref{S:Topoverse},
that {\B} is a complicated 4-manifold {\T}\,---\,the
topoverse\,---\,with no geometry or other decoration.  The ability to
represent {\T} as a labeled graph, with edges corresponding to prime
3-manifolds and vertices to elementary cobordisms, points to
correspondence with Feynman graphs.  This makes the conjecture
plausible.

Highly compressed encoding of accessible features of history as the
mental models of perceptive creatures is central to their
consciousness (at whatever level they function).  It is the
conjunction of time, history and mental models that distinguishes the
representations {\R} from other, uninteresting representations of {\B}.

Acquisition of information and development of mental models is an
interactive process.  We actively participate in our perceived
reality.  We control our bodies and the world we inhabit sufficiently
to collectively investigate, formulate, record, and debate fundamental
ideas regarding the world.  We refine our models by acting on the
world and sensing its response.  Within bounds\,---\,consistent with
our limited present knowledge, a prohibition on revising history, and
severely limited scope and capacity\,---\,we are able to consciously
influence our future.

Contrasting with the perception that we can act independently on the
world is the obvious requirement for mutual consistency of the
perceptions of all observers.  Such consistency can be understood only
if consciousness is actually a single, global
phenomenon.\footnote{Reiner Hedrich has proffered the technical term
  \emph{monistic panpsychism} for this collective consciousness.}  The
perception of individuality is then a robust illusion, just as
quasi-classical particles seem independent even though their apparent
localization relies on their global entanglement with the rest of the
world.

Although I have assumed that {\B} has definite mathematical structure,
it is possible that no mathematical model can fully simulate the real
world.  At a minimum, mathematical formalism must be devised / found
to describe geometry and fields whose uncertainty reflects their
finite information content and the unpredictability of new information
that must accompany progress of time.  Applying such formalism should
then enable resolution of the major puzzles of quantum physics,
general relativity, and their integration.

Reality, as Leibniz insisted, can indeed be found in ``One single
source''.  The ``interconnection of all things with one another''
originates in that source and is maintained in the unity of history,
the alignment of perceptions, and the global integration of classical
and quantum models of the world.

\section{Acknowledgments}

I thank John Moffat for his mentorship and for sharing and debating
diverse ideas on fundamental questions in physics over many years.
Thanks also to the late Robert Brout, who shared his passion for
understanding the foundations of reality, and to Harvey Brown, Olaf
Dreyer, Stephen Green, Reiner Hedrich, Burra Sidharth, Viktor Toth,
and Hans \mbox{Westman} for their perspectives and feedback.

\appendix
\renewcommand{\theequation}{A-\arabic{equation}}
\renewcommand{\alphaeqn}{\setcounter{saveeqn}{\value{equation}}%
\stepcounter{saveeqn}\setcounter{equation}{0}%
\renewcommand{\theequation}
 {\mbox{A-\arabic{saveeqn}\alph{equation}}}}%
\renewcommand{\reseteqn}{\setcounter{equation}{\value{saveeqn}}
\renewcommand{\theequation}{A-\arabic{equation}}}
\setcounter{equation}{0}

\section{Topology as foundation?}\label{S:Topoverse}

The belief that gravity must be a quantum phenomenon, just like matter
fields, led Wheeler to propose that vacuum fluctuations of the
geometry of space could imply topology changes at the Planck scale,
and foam-like paths through superspace \citep{Wheeler57,Wheeler64}.
Exciton-like superpositions involving many different topologies might
be manifested as emergent particles of matter; field lines trapped in
the topology could yield charges \citep{Wheeler68}.  Although the
original proposal involved geometry changes driving topology changes,
Wheeler presented a broader vision of some unknown \emph{pregeometry}
from which the geometrical world would emerge.  The history and
numerous proposals regarding pregeometry are reviewed in
\citep{Gibbs:1995gj}.  Careful analysis of pregeometry proposals
reveals, however, that rather than emerging from some fundamentally
distinct structure, geometry and time have generally been built-in
\citep{Meschini:2004si}.

The bare world, {\B}, of W-1, seems a better fit for Wheeler's
notional pregeometry.  Unlike earlier proposals, the perceived
geometry of space and time are not built-in to {\B}; instead they are
genuinely emergent in \mbox{W-2}.  As first suggested in
\citep{Green80}, a particularly simple proposal for the structure of
\B, inspired by spacetime foam, is as follows:

\medskip\noi\textbf{Conjecture 1 (Topoverse)}: \emph{The bare world,
  {\B}, is a unique, connected, smooth \mbox{4-manifold}, {\T}, with
  complicated global topology but no inherent geometry, fields, or
  other decoration.}

\medskip\noi To emphasize its topological nature and that its
identification with {\B} is only tentative, I shall call {\T} the
\emph{topoverse}.  If the conjecture is true then
\mbox{$\mB\equiv\mT$}.

As the pregeometric foundation, the topoverse has several novel
features: (i)~{\T} is unique, whereas other proposals
(e.g. \citep{Hawking78,Rei6889135}) consider Feynman sums over all or
large classes of 4-manifold topologies; (ii)~the topological
connectivity of {\T} serves as the complete foundation for the real
world throughout time, whereas other proposals assume additional
structure, such as geometry, time, fields, group representations or
causal structure, while generally ignoring the global topological
information; and (iii)~particles and quantum phenomena emerge
naturally from the structure of {\T}, as we shall see below.

\subsection{The topoverse as a labeled graph}\label{SS:Topoverse1}

Points and local neighbourhoods in {\T}, being all equivalent,
contribute no information to the real worlds of W-2.  But the vast
quantity of non-local, relational information residing in the global
structure of {\T} underlies the representations {\R} and their partial
order ``$\prec$\,''.  If we can discover an invariant way of building
up the global structure of {\T} from elementary components then that
will reveal generic properties of its global connectivity that
repeatedly occur in the specific detailed structure of {\T}.  We
approach this by first finding an invariant decomposition of
3-manifolds, and then considering how {\T} can be (quasi-locally)
decomposed as a stack of compact 3-manifolds (and their elementary
cobordisms).

The \emph{connected sum}, $M=M_1 \csum M_2$, of any two $n$-manifolds
is constructed by cutting an open cell out of each of $M_1$ and $M_2$
and identifying the resultant $(n\!-\!1)$-sphere ($S^{n-1}$)
boundaries via a homeomorphism.  The $n$-manifold $M$ is independent
of the choice of open cells.  The connected sum operation is
commutative and associative.

An $n$-manifold is said to be \emph{prime} if $M = M_1 \csum M_2$
implies that at least one of $M_1$, $M_2$ is homeomorphic to $S^n$.
The only compact, prime 2-manifolds are $S^2$, $T^2$ (the torus), and
$P^2$ (the real projective plane).  By contrast, there is a countably
infinity variety of compact, prime 3-manifolds.  Progress on their
classification, in terms of orientability, spinoriality, chirality and
other properties, is reported in \citep{Hatcher_04, Giu4727904}.  The
invariant decomposition of \mbox{3-manifolds} that we need is provided
by the following result:\footnote{We shall be interested only in those
  compact 3-manifolds that are also closed, with no boundary.}

\medskip\noi\textbf{Theorem (3-Manifold Prime Decomposition)}:
\emph{Each connected compact 3-manifold can be uniquely expressed as a
  connected sum of a finite number of prime factors.
  \emph{\citep{Milnor62,Hempel76}}}

\medskip Counter-examples demonstrate that not all 4-manifolds admit a
prime decomposition, although a weaker result has been obtained for
connected compact oriented 4-manifolds \citep{Kr95}.  For that
restricted subset, decomposition is unique only up to connected sum
with any number of closed, simply connected 4-manifolds.  With its
unrestricted, complicated topology, {\T} will, by default, be
non-orientable and have no prime decomposition.  But the construction
below shows how the prime decomposition of 3-manifolds can be
leveraged to represent {\T} as a labeled graph.

Let $g$ be any smooth, positive-definite Riemannian metric on {\T} and
use $g$ to define $d:\mT\!\times\!\mT \rightarrow \mRR$ as the minimum
metrical distance between each pair of points $x,y\in\mT$.  Select an
arbitrary point $O\in\mT$ and define
\alphaeqn%
\beq%
f(x) = d(O, x)%
\eeq%
\beq%
\Sigma_r = \{x\in\mT |\; f(x)=r,\ r>0\}.%
\eeq%
Choose $r_0$ such that $\Sigma_{r_0}$ is a 3-sphere bounding a solid
ball containing $O$.%
\reseteqn%
Because {\T} has complicated global topology, there must be values
$r_i > r_{i-1}$, $i \in \mN$, such that every $\Sigma_{r_i}$ is a
compact 3-manifold but $\Sigma_{r_i}$ is not homeomorphic to
$\Sigma_{r_{i-1}}$.  Adjust $g$, if necessary, so that critical levels
$r$ at which $\Sigma_r$ becomes singular are nondegenerate and $f$ is
a Morse function on {\T} \citep{Milnor69, Guillemin_Pollack74}.
Furthermore, adjust $g$ such that there is only one critical level
$r^c_i$ of $f$ between each $r_i$ and $r_{i-1}$, and all non-essential
critical points are eliminated.  Each region
\beq%
\mT_i = \{x\in\mT |\; r_{i-1}^c+\epsilon < f(x) < r_i^c-\epsilon\},%
\eeq%
where $\epsilon$ is infinitesimal, now has the product topology
$\Sigma_{r_i}\!\times\!\mRR$.

Finally, adjust $g$ so that the prime summands of each $\Sigma_r$, $r$
not a critical level, are separated from each other and confined
within small diameter cylinders in each $\mT_i$ and, where possible,
the cylinders extend smoothly across the critical levels.  At each
critical level the topology change from $\Sigma_{r_{i-1}}$ to
$\Sigma_{r_i}$ will involve only a small number of prime summands, and
the enclosing cylinders for these will naturally meet at the critical
level.  This leaves a geometrical representation of {\T} that has the
form of a graph, $\mG_\mT$.  Edges of $\mG_\mT$ are the above
cylinders, shrunk to infinitesimal radius and labeled according to
the enclosed prime 3-manifold summand.  Nodes or vertices of $\mG_\mT$
are the critical points at which topology changes.

The building blocks of topology change in $\mG_\mT$ are elementary
3-manifold cobordisms involving only a small number of prime
summands.\footnote{Most cobordisms involving $S^3$, which is the
  identity under ``$\csum$'', will be non-essential, and are thus
  eliminated in the construction of $\mG_\mT$ by suitably deforming
  the geometry.}  Selection rules that determine what combinations of
primes may meet at a critical point arise from the simple requirement
that {\T} be a 4-manifold.\footnote{Aspects of this problem are
  addressed in \citep{Ionicioiu97}, with the focus on lower
  dimensions.  Konstantinov \citep{Konstantinov98} applied Morse
  theory in 4-dimensions.}  The abstract graph, with its labeled
edges, provides a representation of {\T}.  The graph, or finite
connected subsets of it, can be embedded in $\mRR^4$; almost every
embedding will be a regular embedding, such that the inclusion map for
each edge is $\textrm{C}^\infty$.  The following seems plausible, and
would lend credibility to Conjecture 1:

\medskip\noi\textbf{Conjecture 2}: \emph{The graph $\mG_\mT$, or some
  reduced graph constructed by grouping the elementary cobordisms into
  equivalence classes, is uniquely determined by {\T}.}\medskip

If, for one or more $i>1$, $\Sigma_{r_i} \equiv S^3$, then $\mG_\mT$
may have multiple disconnected subgraphs.  In that case, our real
world will represent just one highly complicated, connected subgraph
while other subgraphs may correspond to other real worlds.  We shall
not concern ourselves with the possibility of many bare/virtual
worlds, and will thus assume that $\mG_\mT$ has just one connected
component.

Although the above construction was facilitated by an assumed metric,
the graphical representation of {\T} carries only topological
information and will exist independent of any geometry.  The topology
of $\Sigma_r$ can change only in a discrete manner at critical points.
But there is nothing inherent in the topology to allow those critical
points to be localized in {\T}\,---\,interesting topological features are
fundamentally non-local.  Within any connected subgraph, however, one
can define the topological \emph{separation} of any two edges
$\alpha$, $\beta$ as the minimum number of essential cobordisms
between $\alpha$ and $\beta$.

A theorem by Rohlin \citep{Rohlin} establishes that each closed
3-manifold is the boundary of some 4-manifold.  This also follows from
proofs, for both orientable and non-orientable cases, that any closed
3-manifold may be obtained by doing Dehn surgery (see
\citep{Thurston79, Lackenby97}) on a link in either $S^3$ or
$P^2\!\times\!S^1$ \citep{Wallace60, MR0151948, MR0145498}.  A key
implication is that, given any two closed 3-manifolds $S_1$, $S_2$,
the connected sum $M_1 \csum M_2$ of the two 4-manifolds they bound is
a cobordism between them.  Thus all closed 3-manifolds are in just one
cobordism class.  It is important to recognize, however, that even the
simplest cobordism between $S_1$ and $S_2$ may be a composition of
many elementary cobordisms.

\subsection{Quantum matter and geometrical
  spacetime}\label{SS:Topoverse2}

The ability to represent the topoverse as a graph makes plausible the
hypothesis that prime 3-manifolds and their elementary cobordisms (or
equivalence classes thereof) give rise to corresponding real world
matter fields and their interactions.  Distinct matter fields
represent distinct prime 3-manifolds (or equivalence classes);
interaction vertices correspond to the allowed elementary cobordisms
(or equivalence classes).  The discrete properties and interactions of
the various matter fields reflect the generic properties of
4-manifolds, while the specific matter content and history of the
observed universe correspond to the actual detailed structure of {\T}
(and $\mG_\mT$).

With such correspondence, $\mG_\mT$ can be thought of as the complete,
pregeometric, \emph{bare} or \emph{virtual} Feynman graph of the world
or, equivalently, the bare Heisenberg state of the world.  Quantum
field theory and general relativity provide the mathematical
frameworks through which the correspondence is transformed into a
time-space-matter model of the physical aspects of the representations
{\R} of {\T}.  With this model, the mutually-consistent properties of the
\emph{dressed} or \emph{physical} matter we perceive provide, in terms
of geometry and fields on a topologically simple manifold (e.g.
$\mRR^+\otimes\mRR^3$), a very coarse-grained representation of the
complicated global connectivity of {\T}.

Dressed quantum states of identifiable physical subsystems are
projections of $\mG_\mT$ into our present real world, with
observation-based encoding in terms of geometrical spacetime and
physical matter.  But these projections (and their associated Hilbert
space and operators) exclude most of the information in $\mG_\mT$.
The excluded information, since it has not yet been incorporated into
\R, can have no location in our time-space-matter models of {\R}.
Progressive incorporation into {\R}, and our models, of this
previously un-represented information involves consistent updating of
the spacetime geometry and physical matter.  Through this process,
information about {\T}, which might be thought of as \emph{quantum
  information}, becomes represented as part of history and preserved
by causal evolution.  As the states of quantum subsystems are
\emph{reduced}, real world history is enriched.  This, in turn, makes
new dimensions available for the Hilbert space of physical subsystems.

Our models employ spacetime geometry to provide the conceptual (and
mathematical) arena within which quantum matter fields and their
interactions can be placed into sensible correspondence with perceived
reality.  While geometry and matter emerge and evolve in a mutually
consistent manner, all their information content originates from the
topology of {\T} and hence the quantum matter.  Spacetime geometry has
no relevance without matter, and no quantum characteristics beyond
those inherited from matter.

My conjecture that $\mB\equiv\mT$ was largely inspired by the
na\"{i}ve similarity of $\mG_\mT$ to a Feynman graph.  But
classification of prime 3-manifolds, determination of their elementary
cobordisms, and making suitable associations with observed matter
fields remain major challenges for future work to validate the
conjecture.  Whether the conjecture holds up or not, {\T} and $\mG_\mT$
illustrate well the global connectivity, and the kinds of generic and
specific properties, that {\B} must have.

\subsection{Holographic emergence of the real world}\label{SS:Emerge}

The countably infinite variety of distinct prime 3-manifolds and of
elementary 3-manifold cobordisms allows {\T} (and {\B}) to have
sufficient generic and detailed structural information.  Any
observer's evolving real world, $\mR_t$, will encode only a small
fraction of the details.  Like the progressive development of the
image obtained from a hologram, the real world will start as a very
fuzzy representation of {\T} as a whole.  Development of history adds
progressively more information about {\T}, as finer detail, at later
times $t$.

Simplistic geometrical representations of {\T} can be obtained by
mapping the graph $\mG_\mT$ into Minkowski space.  But, due to the
complicated connectivity of $\mG_\mT$, for all such detailed mappings
the images of the graph edges will necessarily have no consistent
relationship to the local null-cone structure; and labels will be
needed to indicate which prime 3-manifold to associate with each edge.

By contrast, our view of the real world at large scale has physical
matter\,---\,radiation, dust, stars, galaxies, clusters\,---\,tracing
out timelike or null geodesic paths in a spacetime whose curvature
reflects the stress-energy-momentum of the matter, in accordance with
the Einstein equations.  At somewhat smaller scales, direct
(non-inertial) interactions\,---\,for example,
electromagnetism\,---\,cause geodesic deviation that can be rationally
explained in terms of intrinsic properties of the matter.

To achieve such a rational relationship between matter and geometry,
our geometry-and-field representation of $\mG_\mT$ must employ a
coarse-grained view, suppressing most details of the bare structure.
The macroscopic geometry of the perceived world then provides a
locally-stable arena within which finer details of $\mG_\mT$ can be
encoded as the observed small-scale states and collective dynamics of
matter.  But these representations will always be limited\,---\,most
details of $\mG_\mT$ will remain at the bare level, collectively
influencing real world history yet resistant to explicit, rational
geometrical representation and thus inaccessible to direct perception.

The interference pattern, recorded on the photographic plate of an
optical hologram, holds the primary information that gives rise to the
the 3-d virtual image perceived by a viewer.  In our proposed
worldview, the bare world $\mB\equiv\mT$ takes the place of the
interference pattern; the virtual holographic image is replaced by the
real world \R.  Holographic emergence of the real world occurs on two
levels.  First is emergence of the universe and its coarse-grained /
large scale geometry, differentiation of matter species, and formation
of structure.  Second is the ongoing development of history through
quantum evolution.  Models for each of these were discussed in Section
\ref{S:Models}.

The first level is like formation of an image by viewing a holographic
plate through a low-pass (i.e. long wavelength) spatial filter (e.g.,
a defocused lens).  The viewer will be unable to discern image details
at scales shorter than the filter's cutoff wavelength.  If that
wavelength is initially comparable to the width of the holographic
plate, then the image will be a formless blur.  Shifting the cut-off
to shorter wavelengths (by gradually focusing the lens) makes the
image progressively sharper and reveals previously unseen structural
details.  Once all details of the interference fringes in the
holographic plate are resolvable, the viewer can develop a more
complete perspective of the 3-d image by varying her vantage point and
the illumination\,---\,this corresponds to the second level of emergence.

The above analogy is useful, but not perfect.  Unlike the external
viewer of a hologram, we construct our collective view of the real
world from within.  Since {\B} has topological information only, not
only does {\R} gain detail as information about {\B} is added, the
entire geometrical time-space-matter mode of representation is
completely emergent.

\begin{flushleft}


\end{flushleft}
\end{document}